# Reconstruction changes drive surface diffusion and determine the flatness of oxide surfaces

Giada Franceschi,[a] Michael Schmid, Ulrike Diebold, and Michele Riva

*Institute of Applied Physics, TU Wien, Wiedner Hauptstraße 8-10/E134, 1040 Wien, Austria*

a) franceschi@iap.tuwien.ac.at

Surface diffusion on metal oxides is key in many areas of materials technology, yet it has been scarcely explored at the atomic scale. This work provides phenomenological insights from scanning tunneling microscopy on the link between surface diffusion, surface atomic structure, and oxygen chemical potential based on three model oxide surfaces: $Fe_2O_3(1\bar{1}02)$, $La_{1-x}Sr_xMnO_3(110)$, and $In_2O_3(111)$. In all instances, changing the oxygen chemical potential used for annealing stabilizes reconstructions of different compositions while promoting the flattening of the surface morphology – a sign of enhanced surface diffusion. It is argued that thermodynamics, rather than kinetics, rules surface diffusion under these conditions: The composition change of the surface reconstructions formed at differently oxidizing conditions drives mass transport across the surface.

## I. INTRODUCTION

Surface diffusion is central to numerous and technologically relevant areas. In (heterogeneous) catalysis, the mobility of reactants on the catalyst's surface determines the course and the rate of the reactions thereby occurring.[1,2] Surface diffusion plays an essential role in strong metal-support interactions[1,3,4] and catalyst deactivation due to sintering of supported metal nanoparticles.[5,6] In metallurgy, surface diffusion determines the sintering rates of metal powders.[7] During the epitaxial growth of thin films (of interest, e.g., for nanoscale integrated devices), the interplay between diffusivity and growth rate of the deposited species determines the growth mode and many film properties.[8]

Metal oxides dominate many technological areas where the role of surface diffusion is vital, e.g., catalysis, thin-film devices, and fuel cells. Yet, the knowledge on surface diffusion on these materials is limited,[9,10] especially when compared to metals and semiconductors.[11-15] This work aims at shedding some light on the processes that govern diffusion on oxide surfaces by using scanning tunneling microscopy (STM) on three different systems: α-$Fe_2O_3$, the most stable iron oxide at ambient conditions,[16] interesting for its potential as a catalyst for photoelectrochemical water splitting;[17,18] Sr-doped $LaMnO_3$ (LSMO), a perovskite oxide with uses and promises for catalytic,[19] energy-conversion[20] and spintronics[21] applications; and $In_2O_3$, a transparent conductive oxide[22,23] used in various catalytic and gas-sensing applications.[24-26] Ultrahigh-vacuum- (UHV-) prepared, single-crystalline $Fe_2O_3(1\bar{1}02)$, $La_{0.8}Sr_{0.2}MnO_3(110)$, and $In_2O_3(111)$ thin films grown by pulsed laser deposition (PLD) were used for the task. The surface of each sample was measured after two annealing treatments at different values of oxygen chemical potential (shortly $\mu_O$), as determined by the chosen temperature and oxygen background pressure (for a definition of $\mu_O$, see the Methods Section). Notably, the temperature was kept constant within each set of experiments (only the $O_2$ pressure was changed), and bulk diffusion and cation evaporation were negligible at the chosen conditions.

The results point to a correlation between atomic structure, oxygen chemical potential, and surface diffusion. Despite the differences in their surface atomic details, all the systems investigated show a consistent behavior. When the annealing conditions promote the formation of a different reconstruction, the surface morphology becomes remarkably flatter. It is argued that the main factor behind the flattening is of thermodynamic nature. When the preparation conditions change, the systems switch between reconstructions to minimize their surface free energy. The different cation composition of the reconstructions creates a strong driving force to displace material across the surface, in turn promoting surface flattening.

## II. EXPERIMENTAL SETUP AND METHODS
### A. *Experimental setup*

All experiments were carried out in a UHV setup that combines a PLD chamber (base pressure $\leq 4 \times 10^{-10}$ mbar after bake-out) with a surface science system (base pressure $\leq 5 \times 10^{-11}$ mbar) through a small interconnected UHV transfer chamber. The PLD apparatus is fitted for high-pressure/high-temperature growth experiments (up to 1200 °C, 1 mbar) while monitoring the growth via reflection high-energy electron diffraction,[27] and was used to grow the films of $In_2O_3(111)$, $Fe_2O_3(1\bar{1}02)$, and LSMO(110) discussed in this work. During growth, a KrF excimer laser (Coherent Compex Pro 201, 248 nm) was directed onto a polished target, and $O_2$ was admitted in the chamber via a leak valve. Substrates were heated by a collimated continuous-wave infrared laser (DILAS Compact Evolution, 980 nm) directed on the back of the samples through a hole in the sample plates, and temperatures were monitored with an Impac IGA5 pyrometer aimed at the sample surface.

Before the growth, the substrates were prepared in an adjoined surface science chamber equipped for $Ar^+$ sputtering and electron-beam annealing, and comprising X-ray photoelectron spectroscopy (XPS, Omicron non-monochromatic dual-anode Mg/Al Kα source, SPECS Phoibos 100 analyzer), low-energy electron diffraction (Omicron SpectaLEED), and scanning tunneling microscopy (SPECS Aarhus 150). The mesoscale morphology of some substrates and films was investigated ex-situ by atomic force microscopy (Agilent 5500 ambient AFM in intermittent contact mode with Si tips on Si cantilevers, both in air and in dry Ar atmosphere). Before the growth, XPS, AFM, and/or STM were used to ensure that the substrates were clean (within the detection limit of XPS) and atomically flat.

To test the dependence of the surface morphology on the annealing conditions, all samples were annealed in the PLD chamber at the conditions specified in the main text (at times following one or more sputtering cycles).

STM images were acquired at room temperature, in constant-current mode, and measuring empty states (positive sample bias $U$). Electrochemically etched W tips were prepared by sputtering, indentation in the film and/or applying voltage/current pulses. STM images were then acquired after moving to a location unaffected by tip conditioning.

## B. *Sample preparation and film growth*

A 0.03 at.% Ti-doped $Fe_2O_3(1\bar{1}02)$ film of 90 nm thickness was grown homoepitaxially on a $(1 \times 1)$-$Fe_2O_3(1\bar{1}02)$ single crystal as detailed in ref. 28. Ti doping is needed to enhance the electrical conductivity of the material. At these concentration levels, it has only a minor influence on the surface structure of the system.[28] The as-received substrate (natural single crystal from SurfaceNet GmbH, $6 \times 6 \times 0.5$ mm$^3$, one-side polished, < 0.3° miscut) was cleaned *in situ* by two cycles of Ar$^+$ sputtering (10 min) plus O$_2$ annealing (1 h, 1 mbar, 900 °C). The film was grown by alternating deposition from $Fe_3O_4$ and 1 at.% Ti-doped $Fe_2O_3$ home-made targets,[29] at 1.7 J/cm$^2$, 5 Hz, $2 \times 10^{-2}$ mbar O$_2$, 850 °C (60 °C/min ramp rates; no post-annealing after growth).

A lanthanum–strontium manganite ($La_{0.8}Sr_{0.2}MnO_3$) film of 22 nm thickness was grown on SrTiO$_3$(110) as detailed in ref. 30. The substrate (single crystal from CrysTec GmbH, 0.5 wt.% Nb-doped, $5 \times 5 \times 0.5$ mm$^3$, one-side polished, miscut < 0.3°) was prepared and characterized in UHV to exhibit a mixture of $(4 \times 1)$ and $(5 \times 1)$ surface reconstructions.[31] The film was grown from a homemade target,[32] at 700 °C, 1 Hz,

1.87 J/cm$^2$, and $1 \times 10^{-2}$ mbar O$_2$. These conditions do not yield stoichiometric transfer from the target to the film and instead cause a significant Mn accumulation at the surface and a rough surface morphology.[30] The surface morphology was partially recovered by 5 cycles of sputtering plus high-O$_2$-pressure annealing (25 min sputtering; annealing for 1 h at 700 °C and 0.2 mbar O$_2$, except for the last cycle that lasted 11 h), which preferentially removed Mn at the surface.

An In$_2$O$_3$(111) film of 600 nm thickness was grown on yttria-stabilized zirconia (YSZ), as detailed in ref. 33. Briefly, the as-received YSZ(111) single crystal (CrysTec GmbH, $5 \times 5 \times 0.5$ mm$^3$, 9.5 mol% Y$_2$O$_3$-doped, one-side polished, $< 0.3°$ miscut) was sonicated in heated neutral detergent (3% Extran MA02, 2× 30 min) and ultrapure water (milli-Q, 10 min), and then annealed in air in a box furnace at 1350 °C for 1 h (ramp rate 8 °C/min; during annealing, the crystal was placed inside pre-annealed and HNO$_3$-cleaned high-alumina crucibles). To ensure sufficient absorption of the infrared laser in the substrate during growth, and to electrically connect the film to its sample plate, the back, the sides, and the four front corners of the otherwise infrared-transparent and insulating YSZ sample were coated with a ~200 nm-thick Ti/Pt bilayer deposited by magnetron sputtering. Following further cleaning (2× 10 min sonication in Extran plus 5 min in ultrapure water), the YSZ sample was mounted on a Nicrofer® (Inconel 602 CA) sample plate with spot-welded Nicrofer clips. After insertion into UHV, the sample was annealed for 20 min in UHV at 550 °C. In$_2$O$_3$ was then grown by PLD from a sintered target (China Rare Metal Material Co., Ltd., 65% density, $\geq 99.99\%$ purity) at 1.7 J/cm$^2$ fluence, 1 Hz repetition frequency, and 0.2 mbar O$_2$ pressure. To achieve a flat and continuous film, a recipe devised in ref. 33 was applied. The substrate temperature was set to 700 °C for the growth of the first 15 nm; the growth was then interrupted, and the sample was annealed for 30 min at 900 °C and 0.2 mbar O$_2$; the growth then proceeded for another 585 nm at 900 °C and 0.2 mbar O$_2$, followed by a 10 min post-annealing at the same conditions.[33]

## C. Definition of oxygen chemical potential

The oxygen chemical potential is defined through the ideal gas expression as $\frac{1}{2}\mu_{O_2}(T, p_{O_2}) = \frac{1}{2}\left[\mu_{O_2}^0(T) + k_B T \ln\left(\frac{p_{O_2}}{p_0}\right)\right]$, where $k_B$ is Boltzmann's constant, $p_0 = 1$ bar, $T$ is the absolute temperature in Kelvin, $\mu_{O_2}^0(T) = H_{O_2}(T, p_0) - H_{O_2}(0, p_0) - T[S_{O_2}(T, p_0) - S_{O_2}(0, p_0)]$ is chosen such that $\mu_{O_2}(0, p_{O_2}) = 0$,[34] and the enthalpy and entropy per $O_2$ molecule, $H_{O_2}(T, p_0)$ and $S_{O_2}(T, p_0)$, respectively, are derived from thermochemical tables.[35]

## III. RESULTS

### A. α-Fe₂O₃(1̄102)

Figure 1 compares the surface morphologies and atomic structures of the same 0.03 at.% Ti-doped Fe₂O₃(1̄102) sample prepared at differently oxidizing treatments. Figure 1(a) shows the surface morphology after one sputtering cycle (15 min, $5 \times 10^{-7}$ mbar Ar⁺, 1 keV, 4.4 µA mm⁻²) plus annealing at mildly oxidizing conditions (30 min, 600 °C, $1.1 \times 10^{-4}$ mbar O₂, or $\mu_O \approx -1.54$ eV; note that the same value of $\mu_O$ is achieved during "standard" surface preparations at 550 °C and $7 \times 10^{-6}$ mbar O₂).[36] As it is common for oxide surfaces after sputtering-annealing treatments, a few atomic layers are exposed, with single-layer-high islands appearing on flat terraces and relatively ragged step edges. The morphology does not significantly improve upon annealing longer at the same conditions. This is expected from the mild dependence of diffusion lengths on annealing duration under kinetically limited conditions.

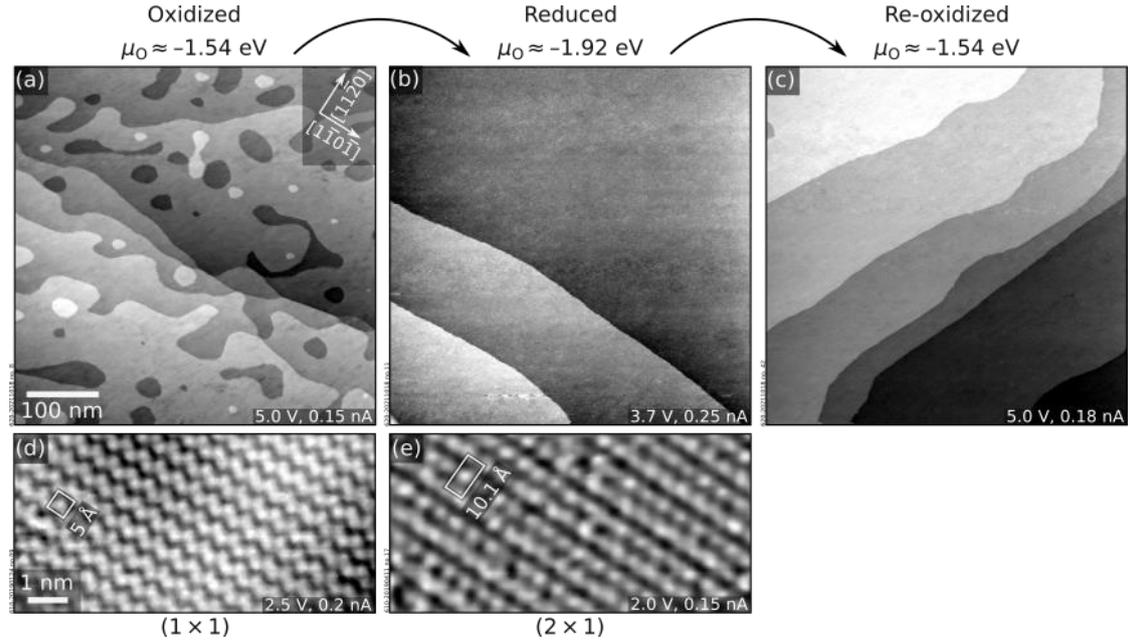

FIG. 1. Surface morphology and atomic structures of $Fe_2O_3(1\bar{1}02)$. (a–c) $500 \times 500$ nm$^2$ STM images of the same Ti-doped $Fe_2O_3(1\bar{1}02)$ film prepared (a) by sputtering plus annealing at oxidizing conditions (30 min, 600 °C, $1.1 \times 10^{-4}$ mbar $O_2$), (b) after subsequent annealing at reducing conditions (30 min, 600 °C, UHV), and (c) after annealing again at oxidizing conditions (30 min, 600 °C, $1.1 \times 10^{-4}$ mbar $O_2$). (d, e) $9 \times 4.5$ nm$^2$ STM images of the oxidized $(1 \times 1)$ (left) and reduced $(2 \times 1)$ (right) terminations observed on the surfaces of panels (a, c) and (b), respectively.

Figure 1(b) shows the appearance of the same sample after subsequent annealing at the same temperature and for the same duration in UHV (pressure during annealing: $5 \times 10^{-9}$ mbar), or at $\mu_O \approx -1.92$ eV. The terraces of the reduced surface drastically increase in size, all single-layer islands disappear, and step edges are now smooth. After annealing the reduced sample at oxidizing conditions once again, the morphology remains flat [Fig. 1(c)].

Figures 1(d) and 1(e) show that the surface morphologies witnessed at differently oxidizing conditions correlate with two distinct, recently investigated[36] atomic structures. Oxidizing conditions stabilize the stoichiometric $(1 \times 1)$ structure of Fig. 1(d), seen as zigzag lines of bright protrusions (Fe atoms) along the $[1\bar{1}0\bar{1}]$ direction, separated by ≈5.0 Å in the $[11\bar{2}0]$ direction. UHV annealing results instead in the reduced $(2 \times 1)$ of Fig. 1(e), showing paired rows of bright protrusions running along the $[1\bar{1}0\bar{1}]$ direction with a 10.1 Å periodicity along the $[11\bar{2}0]$ direction. Recent unpublished findings[37] suggest that the formation of the (2 × 1) structure requires the removal of one Fe and two O atoms per (2 × 1) surface unit cell and a rearrangement of the coordination of Fe atoms in the topmost surface layers. It is possible to reversibly switch between (1 × 1) and (2 × 1) by annealing at differently oxidizing conditions. The surface of Fig. 1(c) exhibits the same (1 × 1) structure as of Fig. 1(a).

## B.  $La_{0.8}Sr_{0.2}MnO_3(110)$

The behavior of LSMO(110) (Fig. 2) is consistent with that of $Fe_2O_3(1\bar{1}02)$. Figure 2(a) shows the surface morphology of an LSMO(110) film after five sputter–annealing cycles at oxidizing conditions (sputtering: 25 min, $5 \times 10^{-6}$ mbar $Ar^+$, 1 keV, 4.4 μA/mm²; annealing: 1 h, 700 °C, 0.2 mbar $O_2$, corresponding to $\mu_O \approx -1.42$ eV). Pits with a depth of a few layers are visible and are caused by non-optimal growth conditions.[30] Several atomic layers are exposed, separated by meandering step edges. The surface morphology improved only marginally after annealing at the same conditions for 11 h.[30,38] Figure 3(b) shows the same surface after annealing for only 1 h at the same temperature but at $7 \times 10^{-6}$ mbar $O_2$ ($\mu_O \approx -1.85$ eV). The number of layers exposed after the reducing treatment decreases significantly, yielding atomically flat terraces with widths of hundreds of nanometers and straight step edges. As in the case of $Fe_2O_3(1\bar{1}02)$, reducing conditions promote surface flattening, and a flat morphology is retained upon annealing back at oxidizing conditions [Fig. 2(c)].

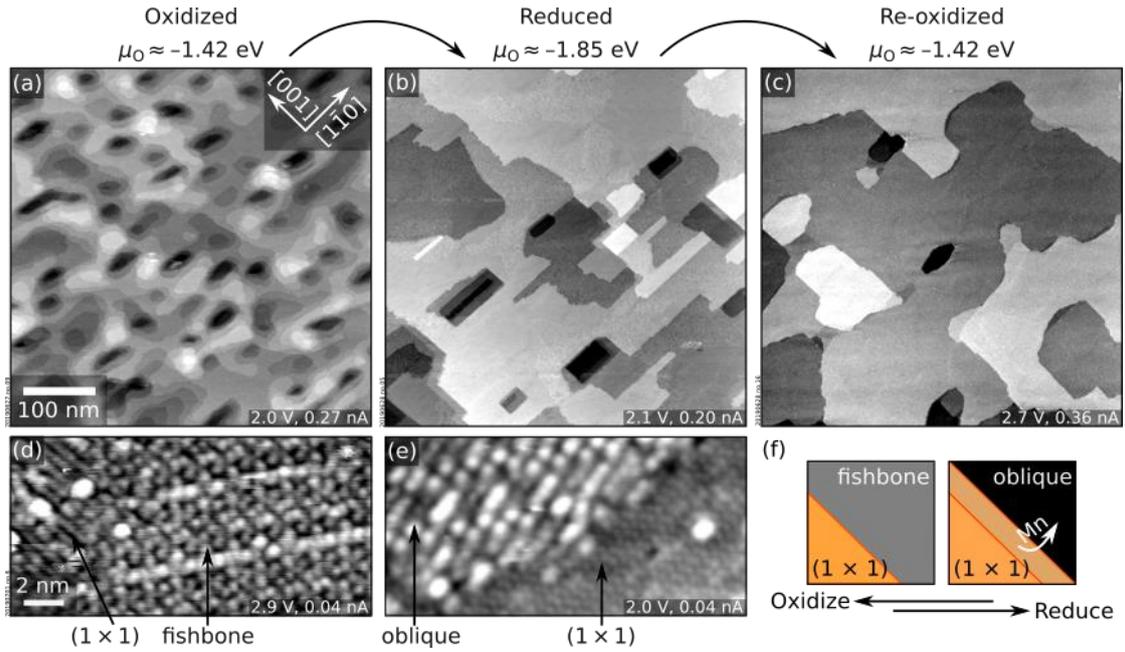

FIG. 2. Surface morphology and atomic structures LSMO(110). (a, b) $500 \times 500$ nm$^2$ STM images of the same LSMO(110) film prepared by (a) 5 cycles of sputtering plus annealing for 1 h at 700 °C and 0.2 mbar O$_2$, followed by (b) annealing for 1 h at 700 °C at $7 \times 10^{-6}$ mbar O$_2$, and (c) annealing again at 700 °C and 0.2 mbar O$_2$. (d, e) $18 \times 9$ nm$^2$ STM images of the oxidized (left) and slightly reduced (right) terminations observed on the surfaces of panels (a) and (b), respectively. Both surfaces exhibit a mixture of different phases. (f) Schematic representation of Mn displacement across the surface occurring at differently oxidizing conditions.

Again, differences are evident at the atomic level [cf. Fig. 2(d) and 2(e)]. As discussed in refs. 32,39, the LSMO(110) surface exhibits various surface phases with different cation and anion compositions, which evolve as a function of $\mu_O$. These include (La, Sr)-rich reconstructions with the $(1 \times 1)$ periodicity of the bulk, and Mn-richer reconstructions with larger unit cells. Annealing over a wide range of $\mu_O$ induces a separation of the surface into (La, Sr)- and Mn-rich phases, while preserving the overall cation concentration.[39] At increasingly reducing conditions, Mn is displaced across the surface to enlarge the $(1 \times 1)$ areas, while the rest of the surface is enriched in Mn, thereby exposing Mn-richer structures [see the sketch in Fig. 2(f)].[39] Comparison of Fig. 2(d) and Fig. 2(e) exemplifies this behavior. Under "oxidizing" conditions [Fig. 2(c)], $(1 \times 1)$ areas are present together with Mn-richer "fishbone" areas.[32] When this surface is annealed at more reducing conditions [Fig. 2(d)], larger $(1 \times 1)$ areas are exposed, while the rest of the surface transitions to the "oblique" structure, richer in Mn by ~1.3 ML than the "fishbone" phase.[39] The process is fully reversible: annealing back at oxidizing conditions produces the high-pressure structures of Fig. 2(c) with the same relative areal coverage.[39]

## C. In$_2$O$_3$(111)

The final example—In$_2$O$_3$(111), Fig. 3—shows consistent behavior with those above. Following preparation at mildly oxidizing conditions (4 min Ar$^+$ sputtering at $1 \times 10^{-7}$ mbar Ar$^+$, 1 keV, 3.4 µA mm$^{-2}$, followed by 30 min annealing at 400 °C at $7 \times 10^{-6}$ mbar O$_2$, or $\mu_O \approx -1.24$ eV), the surface exhibits small and irregular islands of a few nanometers in size [Fig. 3(a)]. After the same sample is reduced by 4 min sputtering at identical conditions plus annealing for 30 min at 400 °C in UHV ($\mu_O \approx -1.47$ eV), the morphology becomes remarkably flatter, exhibiting straight step edges aligned with the low-index directions [Fig. 3(b)]. Morphology flattening under reducing conditions agrees with previous results on the growth of In$_2$O$_3$(111) films at different values of $\mu_O$:[33] Comparatively more reducing conditions yield close-to-equilibrium Stranski-Krastanov growth of 3D islands hundreds of nanometers in size (indicative of high surface diffusion), while more oxidizing conditions result in small and disconnected islands, a sign of kinetically limited growth. Annealing back the reduced sample at oxidizing conditions induces the formation of small islands, as shown in Fig. 3(c). These likely form from the isolated In adatoms of the reduced surface (see below) that react with the O$_2$ supplied by the atmosphere to form In$_2$O$_3$.[40]

Comparing Figs. 3(d) and 3(e) reveals that the changes in the surface morphology are once again accompanied by changes in the surface atomic structure. Under more oxidizing conditions, the surface appears in STM as a threefold-symmetric array of dark, triangular depressions [Fig. 3(d)]. This termination corresponds to a relaxed, bulk-terminated (1 × 1) structure with differently O-coordinated In atoms.[40]

Under "reducing" conditions, bright dots—shown to be In adatoms[40]—appear at the positions of the dark triangles of the oxidized surface [Fig. 3(e)].

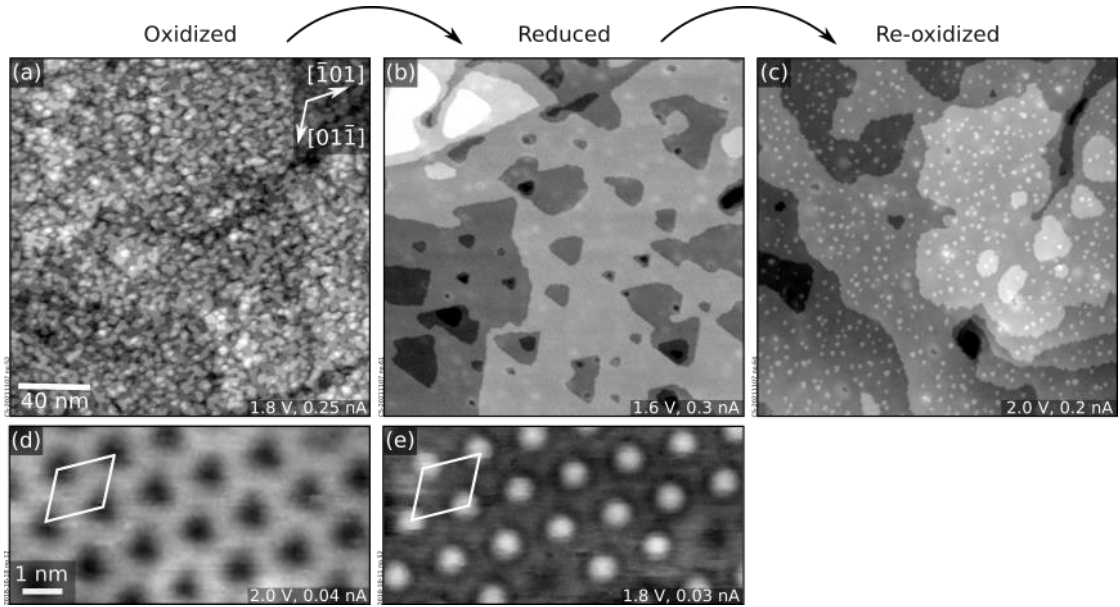

FIG. 3. Surface diffusion and atomic structures of $In_2O_3$(111). (a–c) $200 \times 200$ nm$^2$ STM images of an $In_2O_3$(111) film prepared by subsequent, differently oxidizing treatments: (a) after 4 min sputtering plus annealing for 30 min at 400 °C and $7 \times 10^{-6}$ mbar $O_2$; (b) after 4 min sputtering plus annealing for 30 min at 400 °C in UHV; (c) after annealing for 30 min at 400 °C and $7 \times 10^{-6}$ mbar $O_2$. (d, e) $9 \times 4.5$ nm$^2$ representative STM images of the oxidized and reduced terminations observed on the surfaces of panels (a, c) and (b), respectively. White rhombi highlight the unit cells. Panels (d, e) are adapted with permission from G. Franceschi, M. Wagner, J. Hofinger, T. Krajňák, M. Schmid, U. Diebold, and M. Riva, Phys. Rev. Materials 3, 103403 (2019). Copyright 2019 by the American Physical Society.

Note that, on the epitaxial $In_2O_3$ films used in this work and in contrast to $In_2O_3$ single crystals, the reduced termination cannot be prepared by UHV annealing alone, neither at 400 °C nor at higher temperatures.[33] This is likely because domain boundaries in the film[41] facilitate the otherwise difficult bulk oxygen diffusion from deeper layers to the surface,[42] thereby increasing the "effective" oxygen chemical potential. To obtain the reduced termination, the UHV annealing step must follow a short sputtering cycle (sputtering is preferential to oxygen). Hence, the oxygen chemical potential used to reduce the surface is effectively lower than the value extracted from the UHV annealing parameters alone. Considering that sputtering should promote surface roughening, the flattening of the surface morphology in going from the oxidized to the reduced termination [Fig. 3(a) vs. Fig. 3(b)] is all the more remarkable.

## IV. Discussion

The three case studies presented in this work show consistent behavior. After sputtering plus annealing at oxidizing conditions, the surface morphology is considerably rougher than after a subsequent treatment at more reducing conditions. The flatness is preserved after annealing back at oxidizing conditions. The differently oxidizing treatments concurrently affect the surface atomic details, determining the formation of surface reconstructions with distinct composition: The reduced surface of $Fe_2O_3(1\bar{1}02)$ requires removal of one Fe and two O atoms per $(2 \times 1)$ surface unit cell compared to the oxidized one. At more reducing conditions, the LSMO(110) surface loses O and redistributes its cations across its phase-separated surface. Finally, the reduced $In_2O_3(111)$ surface exposes one additional In adatom per unit cell compared to the oxidized phase. Below it is argued that the change in composition from one reconstruction to another is the main driving force behind the surface flattening observed at differently oxidizing conditions.

Before addressing the mechanisms that drive surface flattening, it is first important to consider the potential role of bulk diffusion and cation evaporation. For some oxides, e.g., $TiO_2$, cation bulk diffusion can be already effective at temperatures as low as 150 °C.[43-45] For the systems considered here, however, the activation barriers for cation bulk diffusion are considerably higher, and bulk cation diffusion is negligible at the conditions employed: For $Fe_2O_3$,[46-48] $In_2O_3$,[41] and LSMO,[49] the cation diffusion lengths derived from the activation barriers available in the literature are around one atomic layer. At the conditions used, one can also discard the role of cation sublimation. Cation evaporation in LSMO at 700 °C was already ruled out in a previous study,[39] while the sublimation of Fe from $Fe_2O_3$ and of In from $In_2O_3$ can be dismissed considering the small vapor pressures of the species (Fe at 600 °C: $1.2 \times 10^{-14}$ mbar;[50] In and In$_2$O at 400 °C: $2.3 \times 10^{-10}$ mbar[51] and $1.12 \times 10^{-9}$ mbar,[52] respectively). Moreover, the formation of loosely bound isolated metal adatoms or clusters is unlikely, based on the large formation enthalpies of metal−oxygen species compared to the values of $\mu_O$ used in this work (FeO: −2.8 eV,[53] MnO: −4.0 eV,[54] In$_2$O$_3$: −3.2 eV per O atom).[55]

Because bulk diffusion and cation evaporation are negligible at the explored annealing conditions, the flattening of the morphologies must be linked to material displacement *across* the surface, in other words, to enhanced surface diffusion. In particular, one expects that the motion of cations or cation–oxygen clusters, rather than oxygen, should be the limiting step for diffusion across the surface. In fact, oxygen can be readily lost/acquired through interaction with the atmosphere. The most natural location for extraction/incorporation of the diffusing species is steps, where undercoordinated sites are present. Consistently, it is common to observe that different types of reconstructions nucleate in the proximity of step edges rather than in the middle of the terraces.

The diffusion of cation species on a given oxide surface is expected to be affected by several factors, including (i) the different chemistry of the various sites present at their reconstructions, precisely the strengths of the metal−oxygen bonds [see the case of LSMO(110)][32] and their basicity [see the case of In$_2$O$_3$(111)],[56] both of which can affect local diffusion barriers;[1] (ii) the density and nature of the diffusing species and point defects (oxygen and cations, cation and oxide clusters, oxygen vacancies, and cation interstitials), whose composition and density can dynamically change through the interaction with the surface and as a function of $\mu_O$; (iii) polarons, i.e., trapped charges coupled to the lattice phonon field expected at transition-metal oxide surfaces.[57] Theory predicted a link between oxygen vacancy migration and polaron hopping in CeO$_2$(111),[58] as well as between polaron formation and adsorption barriers of foreign molecules on rutile TiO$_2$(110),[59] hinting to an influence on self-diffusion barriers. All the factors above may affect the energetic barriers for diffusion and are highly sensitive to the specific atomic details of each surface. It is thus surprising that the systems presented here—characterized by distinct material and surface properties—exhibit similar overall behaviors. This suggests that a more general driving force is responsible for the flattening of the morphologies, which, as explained below, is of thermodynamic nature.

Generally, at equilibrium and at a given value of $\mu_O$, the surface free energy is minimized by stabilizing reconstructions of specific compositions.[14] Consider the case of a surface initially equilibrated at some thermodynamic conditions that stabilize a surface reconstruction of given composition. Afterward, the thermodynamic conditions are changed such that another reconstruction with an excess/deficiency of cations compared to the initial one is more stable. Assuming that bulk diffusion and cation evaporation are negligible, this reconstruction can only be formed if cations are displaced across the surface, taking them from/incorporating them at steps. In other words, when the annealing conditions change such that a different reconstruction must be stabilized, there is an extra thermodynamic driving force for mass transport across the surface. This adds to the kinetic barriers intrinsic to each surface structure that would otherwise determine surface diffusion. Since the systems presented here are characterized by different atomic details, one expects different kinetic barriers for surface diffusion. The fact that all materials exhibit consistent diffusion behaviors suggests that the thermodynamic driving force associated with the change of surface structure composition overrules kinetics, largely determining mass transport, and causing the flattening of the surface morphologies.

It is worth noting that In$_2$O$_3$ makes a partial exception. When annealing back at oxidizing conditions, small islands are formed [Fig. 3(c)], as the In adatoms present under reducing conditions react with oxygen in the atmosphere. The formation of small islands is likely due to an additional diffusion mechanism at play on In$_2$O$_3$, namely the stoichiometry-dependent diffusivity of species formed at differently oxidizing conditions. Growth experiments of $In_2O_3(111)$ at different values of $\mu_O$[33] suggest that the composition of the diffusing species is dictated by $\mu_O$ and may affect their diffusion properties. Specifically, it was found that sub-stoichiometric species such as $In_2O$ form under reducing conditions and that these should diffuse faster than the close-to-stoichiometric $In_2O_3$ species formed under oxidizing conditions. The "slow" $In_2O_3$ species are likely responsible for the formation of the small islands witnessed after re-oxidation.

## V.  CONCLUSIONS AND OUTLOOK

Scanning tunneling microscopy was used to investigate the interplay between surface diffusion, atomic structure, and oxygen chemical potential, $\mu_O$, on three model oxide surfaces, $Fe_2O_3(1\bar{1}02)$, LSMO(110), and $In_2O_3(111)$. In all systems, both the atomic details of the surface and its morphology are affected by the value of $\mu_O$ used to anneal the samples. Improved morphology, i.e., enhanced surface diffusion, is observed when annealing at $\mu_O$ values that stabilize different surface reconstructions than previously present. It was argued that the main driving force behind this effect is the cation composition change in the topmost layers of the surface reconstructions. The composition change adds a thermodynamic driving term for surface diffusion that overrules the kinetic barriers determined by the surface atomic arrangements.

Different metal oxides were investigated: $Fe_2O_3$ as a prototypical binary oxide, LSMO as a complex oxide, and $In_2O_3$ as a post-transition metal oxide. They are characterized by corundum, perovskite, and bixbyite crystal structures, respectively, and exhibit markedly different surface atomic details. The consistent behavior of such a varied spectrum of oxides suggests that it may be shared by many metal oxides. The observed correlation between $\mu_O$ and surface morphology could be used to optimize the preparation of single-crystalline metal-oxide surfaces in UHV: One could enforce surface flattening by alternating annealing at different $\mu_O$s that stabilize reconstructions with different cation compositions. This information could also help optimize oxide film growth. For kinetically limited types of growth, alternating growth at high pressure with annealing at low pressure can help improve the film morphology.

## ACKNOWLEDGMENTS

This work was supported by the Vienna Science and Technology Fund (WWTF), the City of Vienna, and Berndorf Privatstiftung through project MA 16-005. The authors were supported by the Austrian Science Fund (FWF) through project SFB-F81 "Taming Complexity in materials modeling" (TACO) and by European Research Council (ERC) under the European Union's Horizon 2020 research and innovation programme (grant agreement No. 883395, Advanced Research Grant 'WatFun').

## AUTHORS DECLARATIONS

**Conflict of interest**

The authors have no conflicts to disclose.

## DATA AVAILABILITY

The data that support the findings of this study are available from the corresponding author upon reasonable request.